\documentstyle[aps,multicol]{revtex}
\draft
\newcommand{\be}{\begin{equation}}
\newcommand{\ee}{\end{equation}}
\newcommand{\bea}{\begin{eqnarray}}
\newcommand{\eea}{\end{eqnarray}}

\begin{document}
  \begin{flushright} \begin{small}
     DTP-MSU/99-22 \\
     hep-th/9908132
  \end{small} \end{flushright}
\vspace{.5cm}
%%%%  Title  %%%%
\begin{center}
{\large \bf Intersecting $M$--Fluxbranes}
\vskip.5cm
%%%%%  Authors  %%%%
Chiang-Mei Chen\footnote{Email: chen@grg1.phys.msu.su},
Dmitri V. Gal'tsov\footnote{Email: galtsov@grg.phys.msu.su}
and Sergei A. Sharakin\footnote{Email: sharakin@grg1.phys.msu.su}
\vskip.2cm
%%%%%  Address  %%%%
\smallskip
{\em Department of Theoretical Physics,
     Moscow State University, 119899, Moscow, Russia}
\vskip.1cm
%%%%%  Date  %%%%
%{\small(March 18, 1999)}
\end{center}
%\maketitle

\begin{abstract}
New solution to the six--dimensional vacuum Einstein's equations is
constructed as a non-linear superposition of two five-dimensional solutions
representing the Melvin--Gibbons--Maeda Universe and its $S$-dual.
Then using duality between $D=8$ vacuum and a certain class of $D=11$
supergravity configurations we generate $M2$ and $M5$ fluxbranes as well
as some of their intersections also including waves and KK-monopoles.
\end{abstract}

%\pacs{PACS number(s): 04.20.Jb, 04.50.+h, 46.70.Hg}
\begin{multicols}{2}
\narrowtext

%%%%%%%%%%%%%%%%%%%%%%%%%%%%%%%%%%%%%%%%%%%%%%%%%%%%%%%%%%%%%%%%%%%%%%
%%%%    Introduction                                              %%%%
%%%%%%%%%%%%%%%%%%%%%%%%%%%%%%%%%%%%%%%%%%%%%%%%%%%%%%%%%%%%%%%%%%%%%%
\section{Introduction}
As it is well-known \cite{Gi86}
when gravity is taken into account, an uniform
electromagnetic field gives rise to spacetimes of two different types.
One is the Bertotti-Robinson solution which is the product
$AdS_2\times S^2$ and thus is $SO(3)$ symmetric, the field is
`radially' directed. The second is the Melvin Universe \cite{Me64}, which
is the static cylindrically symmetric solution to the four-dimensional
Einstein-Maxwell theory, the field being directed along the axis of symmetry.
In the context of the four-dimensional $N=2$
supergravity the first solution is fully supersymmetric, while the second is
not. It is, however, of interest in view of the possibility
to study non-perturbative processes of  pair creation  of charged
black holes in strong electromagnetic field \cite{DoGaGiHo94}.
Similar situation holds in $D=11$ supergravity where instead of
the two-form field one encounters a four-form. Correspondingly,
there exists a fully supersymmetric solution $AdS_4\times S^7$,
analogous to the Bertotti-Robinson solution of the four-dimensional
Einstein-Maxwell theory, while the Melvin type gives rise to
so-called fluxbranes, term suggested in \cite{DoGaGiHo96}. Simplest
explicit fluxbrane solution for higher rank antisymmetric form
was obtained in \cite{GaRy98}
using some generating symmetry. Here we construct more general fluxbrane
solutions to $D=11$ supergravity including intersecting ones employing
another technique.

%%%%%%%%%%%%%%%%%%%%%%%%%%%%%%%%%%%%%%%%%%%%%%%%%%%%%%%%%%%%%%%%%%%%%%
%%%%                                                              %%%%
%%%%%%%%%%%%%%%%%%%%%%%%%%%%%%%%%%%%%%%%%%%%%%%%%%%%%%%%%%%%%%%%%%%%%%
\section{Six-Dimensional Solution}
The original Melvin solution \cite{Me64} representing the gravitational
field of a magnetic flux tube in cylindrical coordinates reads
 \bea
ds_4^2 &=& U^2 ( - dt^2 + d\chi^2 + d\rho^2)
    + \frac{\rho^2}{U^2} d\phi^2, \nonumber \\
A_\phi &=& -\frac{a}{U}, \qquad U = 1 + \frac{\rho^2}{a^2},  \label{mel}
\eea
where $a= 2/B,\, B$ being the magnetic field strength.
Magnetic field is confined within the tube of radius $a$, so that the
total flux is equal to $B^{-1}$. The transversal space has an infinite
volume, but the circles of constant $\rho$ have the circumferences going
to zero at infinity. The same metric describes an electric Melvin solution
in view of the continuous electric-magnetic duality of the Einstein-Maxwell
system.

If one wishes to consider the Maxwell field as originating from the fifth
dimension via Kaluza-Klein reduction, one has to use the following
parameterization of the five-dimensional line element
\be
ds_5^2 = e^{-\frac4{\sqrt{3}}\phi} (dx_5 + 2 A_\mu dx^\mu)^2
       + e^{\frac2{\sqrt{3}}\phi} g_{\mu\nu} dx^\mu dx^\nu,
\ee
which also contains a dilaton in four dimensions, as it is clear from
the reduction of the five--dimensional Einstein--Hilbert action
(up to surface terms)
\be  \label{ac}
S = \int d^4x \sqrt{-g} \left\{ R - 2(\partial\phi)^2
  + e^{-2{\sqrt{3}}\phi} F^2 \right\},
\ee
with $F=d A$.

The generalization of the Melvin solution (\ref{mel}) to
Kaluza--Klein theory was constructed by Gibbons and Maeda \cite{GiMa88},
in terms of the four--dimensional fields it reads
\bea
ds_4^2 &=& U^\frac12 ( - dt^2 + d\chi^2 + d\rho^2)
    + U^{-\frac12} \rho^2 d\phi^2, \nonumber \\
e^{-\frac4{\sqrt{3}}\phi} &=& U, \quad
  A_\phi = \frac{\rho^2}{2aU}, \quad
    \label{GM}
\eea
Alternatively, this field configuration represents the solution of
the five-dimensional vacuum Einstein's equations as follows
\bea
ds_5^2 &=& U \left( dz + \frac{\rho^2}{aU} d\phi \right)^2
  \nonumber \\
  &-& dt^2 + d\chi^2 + d\rho^2 + \frac{\rho^2}{U} d\phi^2. \label{GM1}
\eea

In presence of two Killing vectors $\partial_t, \partial_5$ the action
(\ref{ac}) exhibits a discrete S-duality under interchange of electric
and magnetic sectors with the simultaneous change of sign of the dilaton.
That way one obtains the following solution
\bea
ds_5^2 &=& V^{-1} ( dy + 2 b^{-1}\chi dt )^2 \nonumber \\
   &+& V ( - dt^2 + d\chi^2 + d\rho^2) + \rho^2 d\phi^2, \label{BR1} \\
V &=& 1 + \frac{\rho^2}{b^2}. \nonumber
\eea
In terms of four--dimensional fields, it reads
\bea
ds_4^2 &=& V^\frac12 ( - dt^2 + d\chi^2 + d\rho^2)
    + V^{-\frac12} \rho^2 d\phi^2, \nonumber \\
e^{-\frac4{\sqrt{3}}\phi} &=& V^{-1}, \quad
  A_t = \chi/b. \quad \label{BR}
\eea

Remarkably, these two five-dimensional vacuum solutions (\ref{GM},\ref{BR})
can be combined
through a direct ``superposition'', into the following six-dimensional
vacuum solution
\bea
ds_6^2 &=& V^{-1} ( dy + 2 b^{-1} \chi dt )^2
    + U \left( dz + \frac{\rho^2}{aU} d\phi \right)^2
    \nonumber \\
   &+& V ( - dt^2 + d\chi^2 + d\rho^2) + \frac{\rho^2}{U} d\phi^2.
   \label{6DM}
\eea
This solution will be used to generate new classes
of $11D$ supergravity configurations via the following non-locally realized
symmetry.

%%%%%%%%%%%%%%%%%%%%%%%%%%%%%%%%%%%%%%%%%%%%%%%%%%%%%%%%%%%%%%%%%%%%%%
%%%%                                                              %%%%
%%%%%%%%%%%%%%%%%%%%%%%%%%%%%%%%%%%%%%%%%%%%%%%%%%%%%%%%%%%%%%%%%%%%%%
\section{$D=8$ vacuum -- $D=11$ supergravity correspondence}
One can show \cite{ChGaSh99} that there exist a ``duality'' between $8D$
vacuum configurations
possessing two commuting space--like Killing vectors and $D=11$ supergravity
solutions satisfying a certain ans\"atz.
The eight-dimensional metric possessing two commuting Killing vectors
in suitable coordinates reads
\bea
ds_8^2 &=& h_{ab} (dz^a+A_{[1]\mu}^a dx^\mu)(dz^b+A_{[1]\nu}^b dx^\nu)
    \nonumber \\
    &+& (\det h)^{-\frac14} g_{\mu\nu} dx^\mu dx^\nu,
\eea
where the $2 \times 2$ real matrix $h_{ab}$ and the one-forms $A_{[1]\mu}^a$
depend only on $x^\mu$. For the $11D$ metric one assumes the following
ans\"atz:
\bea
ds_{11}^2 &=& g_2^\frac12 \delta_{ab} dz^a dz^b
    + g_3^\frac13 \delta_{ij}  dy^i dy^j \nonumber \\
   &+& g_2^{-\frac14} g_3^{-\frac14} g_{\mu\nu} dx^\mu dx^\nu, \label{MA}
\eea
where the functions $g_2, g_3$ and the six-dimensional metric $g_{\mu\nu}$
depend only on {\em transverse space--time} coordinates $x^\mu$, while
for the  four-form  the corresponding decomposition should hold
\bea
&\hat F_{[4]\mu\nu ab} = \epsilon_{ab} F_{[2]\mu\nu},& \qquad
\hat F_{[4]\mu ijk} = \epsilon_{ijk} \partial_\mu \kappa, \nonumber \\
&\hat F_{[4]\mu_1\mu_2\mu_3\mu_4} = H_{[4]\mu_1\mu_2\mu_3\mu_4},& \label{FA}
\eea

The above duality goes as follows. One has to build the  matrix $h_{ab}$
and construct two-forms from KK-potentials
\bea
h_{ab} &=& e^{\frac23\psi} \left( \begin{array}{cc}
                  e^{-\phi}+\kappa^2 e^\phi & \kappa e^\phi \\
                  \kappa e^\phi & e^\phi
                  \end{array} \right), \nonumber \\
F_{[2]} &=& d A_{[1]}^1, \qquad H_{[2]} = d A_{[1]}^2,
    \label{DT}
\eea
where $\phi$ and $\psi$ are defined by
\be
\phi = -\frac12 \ln g_3, \qquad
\psi = -\frac34 \ln g_2 - \frac14 \ln g_3, \label{Phi}
\ee
and $H_{[2]}$ is a dual of the four--form field $H_{[4]}$
in the transverse space--time defined by
\be
H_{[4]\alpha_1\cdots\alpha_4} = \frac12 \sqrt{-g} e^{\psi+\phi}
  \epsilon_{\alpha_1\cdots\alpha_4\mu\nu}
  \left( H_{[2]}^{\mu\nu} + \kappa F_{[2]}^{\mu\nu} \right), \label{H2}
\ee

It has to be noted that
the mapping from $8D$ to $11D$ is a one to many,
i.e. an $8D$ vacuum solution can have several distinct $11D$ supergravity
counterparts
depending on the choice and the order of two Killing vectors in eight
dimensions.

%%%%%%%%%%%%%%%%%%%%%%%%%%%%%%%%%%%%%%%%%%%%%%%%%%%%%%%%%%%%%%%%%%%%%%
%%%%                                                              %%%%
%%%%%%%%%%%%%%%%%%%%%%%%%%%%%%%%%%%%%%%%%%%%%%%%%%%%%%%%%%%%%%%%%%%%%%
\section{Single $M$--Fluxbranes}
To use the above procedure for the solution (\ref{6DM}) one has to
add two extra coordinates $x_1, x_2$ so that $ds_8^2=ds_6^2+dx_1^2+dx_2^2$
and then to choose the pair of Killing vectors with respect to which
the parameterization of the eight-dimensional metric has been done.
Let us start with deriving single $M2$-- and $M5$--fluxbranes
by uplifting either (\ref{GM}) or (\ref{BR}) $5D$ solutions translated
to $8D$ form by adding three flat space coordinates.
Uplifting of (\ref{GM}) with respect to the ordered pair of the Killing
vectors $(\partial_z,\partial_x)$ ($x$ can be any one of the added
three flat coordinates)
one obtains the $M2$--fluxbrane
\bea
ds_{11}^2 &=& U^{-\frac23} \Big\{\rho^2 d\phi^2 + dy_1^2 + dy_2^2 \Big\}
    \nonumber \\
   &+& U^\frac13 \Big\{ - dt^2 + d\chi^2 + d\rho^2
    + dy_3^2 + \cdots + dy_7^2 \Big\}, \nonumber \\
\hat A_{\phi 12} &=& \frac{\rho^2}{aU}. \label{M2}
\eea
Alternatively, one can get the same solution by uplifting
(\ref{BR}) with respect to the Killing pair
$(\partial_x,\partial_y)$, this is a more complicated way.
Note that contrary to the canonical charged $M2$--brane,
the $M2$--fluxbrane corresponds not to electric, but to the magnetic
sector of the form field.

Similarly, the $M5$--fluxbrane can be derived by uplifting (\ref{BR})
using the following ordered pair of Killing vectors
$(\partial_y,\partial_x)$ (simpler way) or uplifting (\ref{GM}) via
$(\partial_x,\partial_z)$:
\bea
ds_{11}^2 &=& V^{-\frac13} \Big\{\rho^2 d\phi^2
    + dy_1^2 \cdots + dy_5^2 \Big\} \nonumber \\
   &+& V^\frac23 \Big\{ - dt^2 + d\chi^2 + d\rho^2
    + dy_6^2 + dy_7^2 \Big\}, \nonumber \\
\hat A_{t67} &=& 2 b^{-1} \chi. \label{M5}
\eea
This solution is supported by the electric sector of the four-form,
contrary to the standard solitonic five-brane.
Both solutions does not possess residual supersymmetry.

%%%%%%%%%%%%%%%%%%%%%%%%%%%%%%%%%%%%%%%%%%%%%%%%%%%%%%%%%%%%%%%%%%%%%%
%%%%                                                              %%%%
%%%%%%%%%%%%%%%%%%%%%%%%%%%%%%%%%%%%%%%%%%%%%%%%%%%%%%%%%%%%%%%%%%%%%%
\section{Intersecting $M$--Fluxbranes}
Now let us derive the $11D$ counterparts to the six-dimensional
solution (\ref{6DM}) translated in two flat directions, $x_1, x_2$,
making different choice of the ordered pairs from three Killing vectors
$\partial_y, \partial_z$
and $\partial_x$ ($x$ could be $x_1$ or $x_2$).
They form a class $11D$ supergravity solutions of the intersecting
fluxbrane type.  Different choice of Killing vectors leads to the
following intersection:

\begin{center}
\begin{tabular}{|p{1.5cm}p{1.5cm}@{}p{5cm}|} \hline
  \multicolumn{2}{|c}{Killing Vectors}& $11D$ Solution \\ \hline
  $\hspace{15pt}\partial_z$ & $\partial_y$ & $M2 \bot M2$-fluxbranes \\
  \hline
  $\hspace{15pt}\partial_y$ & $\partial_z$ & $M5 \bot M5$-fluxbranes \\
  \hline
  $\hspace{15pt}\partial_z$ & $\partial_x$ & $M2$-fluxbrane $\bot$ Wave \\
  \hline
  $\hspace{15pt}\partial_x$ & $\partial_z$ & $M5$-fluxbrane $\bot$ Wave \\
  \hline
  $\hspace{15pt}\partial_x$ & $\partial_y$ & $M2$-fluxbrane $\bot$ Monopole \\
  \hline
  $\hspace{15pt}\partial_y$ & $\partial_x$ & $M5$-fluxbrane $\bot$ Monopole \\
  \hline
\end{tabular}
\end{center}

First let us consider the case of the pair of Killing vectors
$(\partial_z,\partial_y)$ in some detail.
Following (\ref{DT}), the transition to eleven dimension goes through
the quantities
\be
\phi = \frac12 \ln\left( V^{-1} U^{-1} \right), \qquad
\psi = \frac34 \ln\left( V^{-1} U \right),
\ee
>from which the functions $g_2$ and $g_3$ can be obtained by solving
(\ref{Phi}):
\be
g_2 =V^{\frac23} U^{-\frac43}, \qquad g_3 =V U.
\ee
Then the metric of the corresponding $11D$-supergravity solution
easily reads from (\ref{MA})
\bea
ds_{11}^2 &=& V^{-\frac23} U^{-\frac23} \rho^2 d\phi^2
    + V^\frac13 U^{-\frac23} \Big\{ dy_1^2 + dy_2^2 \Big\}
    \nonumber \\
   &+& V^{-\frac23} U^\frac13 \Big\{ dy_6^2 + dy_7^2 \Big\}
    \nonumber \\
   &+& V^\frac13 U^\frac13 \Big\{ \!-\!dt^2\!+\!d\chi^2\!+\!d\rho^2
   \!+\!dy_3^2\!+\!dy_4^2\!+\!dy_5^2 \Big\},
\eea
while the potential of $11D$ four--form field can be found
according to (\ref{FA},\ref{H2})
\be
\hat A_{\phi 12} = \frac{\rho^2}{aU}, \qquad
\hat A_{\phi 67} = \frac{\rho^2}{bV}.
\ee
The solution obtained is the intersection of two $M2$--fluxbranes.
For the case $a^{-1}=0$ (or $b^{-1}=0$)  it reduces to
usual $M2$--fluxbrane located at $(\phi, y_6, y_7)$ (or $(\phi, y_1, y_2)$).

Different combination of pair Killing vectors generates through similar
calculations other $11D$ solutions which we present below.
The pair $(\partial_y,\partial_z)$ generates intersecting of two
$M5$--fluxbranes:
\bea
ds_{11}^2 &=& V^{-\frac13} U^{-\frac13} \Big\{ \rho^2 d\phi^2
    + dy_3^2 + dy_4^2 + dy_5^2 \Big\} \nonumber \\
   &+& V^\frac23 U^{-\frac13} \Big\{ dy_1^2 + dy_2^2 \Big\}
    + V^{-\frac13} U^\frac23 \Big\{ dy_6^2 + dy_7^2 \Big\}
    \nonumber \\
   &+& V^\frac23 U^\frac23 \Big\{ - dt^2 + d\chi^2 + d\rho^2 \Big\},
    \nonumber \\
\hat A_{t12} &=& 2 b^{-1} \chi, \qquad
\hat A_{t67} = 2 a^{-1} \chi.
\eea

The pair $(\partial_z,\partial_x)$ gives the superposition of
$M2$--fluxbrane with wave--type solution
\bea
ds_{11}^2 &=& U^{-\frac23} \Big\{\rho^2 d\phi^2 + dy_1^2 + dy_2^2 \Big\}
    \nonumber \\
   &+& U^\frac13 \Big\{ V^{-1} ( dy + 2b^{-1}\chi dt )^2 \nonumber \\
   &+& V ( - dt^2 + d\chi^2 + d\rho^2 )
    + dy_3^2 + \cdots + dy_6^2 \Big\}, \nonumber \\
\hat A_{\phi 12} &=& \frac{\rho^2}{aU},
\eea
while the pair $(\partial_x,\partial_z)$ generates the superposition
of $M5$--fluxbrane and wave:
\bea
ds_{11}^2 &=& U^{-\frac13} \Big\{\rho^2 d\phi^2
    + dy_1^2 \cdots + dy_5^2 \Big\} \nonumber \\
   &+& U^\frac23 \Big\{ V^{-1} ( dy + 2b^{-1}\chi dt )^2 \nonumber \\
   &+& V ( - dt^2 + d\chi^2 + d\rho^2 ) + dy_6^2 \Big\}, \nonumber \\
\hat A_{ty6} &=& 2 a^{-1} \chi.
\eea

The remaining pair $(\partial_x,\partial_y)$ generates a superposition
of $M2$--fluxbrane with monopole--type solution
\bea
ds_{11}^2 &=& V^{-\frac23} \Big\{\frac{\rho^2}{U} d\phi^2
    + U ( dz + \frac{\rho^2}{aU}d\phi )^2
    + dy_6^2 \Big\} \nonumber \\
   &+& V^\frac13 \Big\{ - dt^2 + d\chi^2 + d\rho^2
    + dy_1^2 \cdots + dy_5^2 \Big\}, \nonumber \\
\hat A_{\phi z6} &=& \frac{\rho^2}{bV}.
\eea
while the reordered pair $(\partial_y,\partial_x)$ gives a superposition
of $M5$--fluxbrane and monopole:
\bea
ds_{11}^2 \!&=&\! V^{-\frac13} \Big\{\frac{\rho^2}{U} d\phi^2
  \!+\! U ( dz\!+\!\frac{\rho^2}{aU}d\phi )^2
  \!+\! dy_3^2\!+\!\cdots\!+\!dy_6^2 \Big\} \nonumber \\
   &+& V^\frac23 \Big\{ - dt^2 + d\chi^2 + d\rho^2
    + dy_1^2 + dy_2^2 \Big\}, \nonumber \\
\hat A_{t12} &=& 2 b^{-1} \chi.
\eea

%%CCM
\section{Discussion}
Thus in eleven-dimensional supergravity there exist a variety
of fluxbrane solutions originating from $M2$ and $M5$ fluxbranes.
Contrary to odinary charged branes, the $M2$ fluxbrane is magnetic, while
the $M5$ one -- electric. Their intersections obey the same intersection
rules as those for ordinary $p$-branes. Our class does not
include intersections of $M2$ and $M5$ fluxbranes between themselves,
but these are likely to exist too as well as intersections of more
than two branes. All these solutions are non supersymmetric. Using
similar method one can construct more general solutions including
combinations of fluxbranes and ordinary charges branes and investigate
the creation of the latters in the field of external four-forms.

%\section*{Acknowledgments}
%\newpage
%%%%%%%%%%%%%%%%%%%%%%%%%%%%%%%%%%%%%%%%%%%%%%%%%%%%%%%%%%%%%%%%%%%%%%
%%%%    References                                                 %%%%
%%%%%%%%%%%%%%%%%%%%%%%%%%%%%%%%%%%%%%%%%%%%%%%%%%%%%%%%%%%%%%%%%%%%%%

\end{multicols}
\end{document}